\documentclass[galaxies,article,accept,pdftex,moreauthor]{Definitions/mdpi} 

\firstpage{1} 
\pubvolume{1}
\issuenum{1}
\articlenumber{0}
\pubyear{2023}
\copyrightyear{2023}
\externaleditor{Academic Editor: Maria Giovanna Dainotti}
\datereceived{30 September 2023  } 
\daterevised{4 November 2023  } % Comment out if no revised date
\dateaccepted{14 November 2023  } 
\datepublished{ } 
\hreflink{https://doi.org/} % If needed use \linebreak
\Title{Holographic Quantum-Foam Blurring, and Localization of Gamma-Ray Burst GRB221009A}

\TitleCitation{Holographic Quantum-Foam Blurring, and Localization of Gamma-Ray Burst GRB221009A}

\Author{Eric Steinbring} %MDPI: Please carefully check the accuracy of names and affiliations.

\AuthorNames{Eric Steinbring}

\AuthorCitation{Steinbring, E.}
\address [1] {%
National Research Council Canada, Herzberg Astronomy and Astrophysics, Victoria, BC V9E 2E7, Canada; eric.steinbring@nrc-cnrc.gc.ca %MDPI: We added the email addresses here according to those submitted online at susy.mdpi.com. Please confirm.
}

\abstract{Gamma-ray burst GRB221009A was of unprecedented brightness in the $\gamma$-rays and X-rays through to the far ultraviolet, allowing for  identification within a host galaxy at redshift $z=0.151$ by multiple space and ground-based optical/near-infrared telescopes and enabling a first association---via cosmic-ray air-shower events---with a photon of 251 TeV. That is in direct tension with a potentially observable phenomenon of quantum gravity (QG), where spacetime ``foaminess” accumulates in wavefronts propagating cosmological distances, and at high-enough energy could render distant yet bright pointlike objects invisible, by effectively spreading their photons out over the whole sky. But this effect would not result in photon loss, so it remains distinct from any absorption by extragalactic background light. A simple multiwavelength average of foam-induced blurring is described, analogous to atmospheric seeing from the ground. When scaled within the fields of view for the {\emph{Fermi}} %MDPI: Please confirm if the italics is unnecessary and can be removed. The following highlights are the same. ***Author response: I assumed, based on some styles used elsewhere, that spacecraft names are always italicized; if MDPI does not, that is fine, please go ahead and remove italics for all spacecraft***
 and {\emph{Swift}} instruments, it fits all $z\leq 5$ GRB angular-resolution data of 10 MeV or any lesser peak energy and can still be consistent with the highest-energy localization of GRB221009A: a limiting bound of about 1 degree is in agreement with a holographic QG-favored formulation.}

\keyword{{gravitation; gamma rays; bursts} %MDPI: please check if this is only one keyword, we suggest using following format: keyword 1; keyword 2; keyword 3 (List three to ten pertinent keywords specific to the article yet reasonably common within the subject discipline). %please confirm intended meaning has been retained ***Author response: This is fine as you have it ***
} 

\begin{document}

%%%%%%%%%%%%%%%%%%%%%%%%%%%%%%%%%%%%%%%%%%
%\setcounter{section}{-1} %% Remove this when starting to work on the template.
%\section{How to Use this Template}
%
%The template details the sections that can be used in a manuscript. Note that the order and names of article sections may differ from the requirements of the journal (e.g., the positioning of the Materials and Methods section). Please check the instructions on the authors' page of the journal to verify the correct order and names. For any questions, please contact the editorial office of the journal or support@mdpi.com. For LaTeX-related questions please contact latex@mdpi.com.%\endnote{This is an endnote.} % To use endnotes, please un-comment \printendnotes below (before References). Only journal Laws uses \footnote.

% The order of the section titles is different for some journals. Please refer to the "Instructions for Authors” on the journal homepage.

\section{Introduction}\label{introduction}

Although general relativity (GR) neglects the quantum nature of particles, requiring a smooth metric, a successful theory of quantum gravity (QG) must account for inherent uncertainty as energies, lengths and timescales approach the Planck scale: an irreducible ``foamy” microscopic spacetime structure, as first proposed by {Wheeler} \citep{Wheeler1957}. %MDPI: Please cite all references with reference numbers and place the numbers in square brackets ("[ ]"), e.g., [1], [1-3], or [1,3]. Please refer to the following website for more information: http://www.mdpi.com/authors/references.  we deleted [1957; see][for a recent review], please confirm, same with other references format; 
%References should be numbered in order of appearance. We detected "Ref 6" appears before "1-5", and refs. [1,22,33] are note cited in maintext, please add citation, please rearrange all the references to appear in numerical order.
See \citep{Carlip2023} for a recent review. %[1957; see][for a recent review]. ***Author: Okay, all references are corrected for style; re-ordered, numbered by order of appearance.
One phenomenology that may probe this is whether tiny, continual, random distance fluctuations $\pm \delta l$ proportional to the Planck length $l_{\rm P} \sim {10}^{-35}$ m (or, equivalently, the timescale $t_{\rm P}\sim10^{-44}~{\rm s}$) accumulate in electromagnetic wavefronts, as they travel long distances through the spacetime foam. The strength of their phase degradation at the observed wavelength $\lambda$ would depend on the summation of the phase perturbations $\Delta \phi = 2\pi \delta l/\lambda$ along the trajectory of the length $L$: $$\Delta\phi_0=2\pi a_0 {l_{\rm P}^{\alpha}\over{\lambda}}L^{1 - \alpha}\eqno(1)$$%please confirm intended meaning has been retained ***Author: Fine.***
where for $a_0\sim 1$ and $\alpha$ specifying the quantum gravity (QG) model, $1/2$ implies a random walk and $2/3$ is consistent with the holographic principle; it vanishes for $\alpha=1$ \citep{Lieu2003}. If the effect is present, distant pointlike objects may appear blurry, although in optical light this could only be comparably so to the diffraction limit of the {\emph{Hubble} \emph{Space } \emph{Telescope}}. But that already strongly rules out the random walk case ($\alpha=0.5$) and constrains blurring to be weak ($\alpha\geq 0.65$) %***Author: The equality sign was the wrong way, so I flipped it over; now fine.***
by images of distant galaxies \citep{Ng2003, Ragazzoni2003} and active galactic nuclei (AGNs)~\citep{Steinbring2007, Christiansen2011, Perlman2011, Tamburini2011}. Another promising method is to use high-resolution spectroscopy and look for the spread of narrow emission lines in distant galaxies, which sets a similar limit \citep{Cooke2020}.

Within distant galaxies, gamma-ray bursts (GRBs) provide another useful target to probe the QG regime by looking for Lorentz invariance, due to the long distance over which their high-energy $\gamma$-rays travel. As energies approach the Planck energy \mbox{$E_{\rm P}\sim 10^{28}~{\rm eV}$,} fluctuations could scale linearly with dispersion $\delta E$ as $L/c$ \citep{Amelino-Camelia1998}; so far, %please confirm intended meaning has been retained ***Author: Fine.***
they have been found to be inconsistent with QG formulations falling outside a lower limit of 1.2 $ E_{\rm P}$ \citep{Liu2022}. But GRBs could instead place strict observational limits on foam-induced blurring by virtue of being nearly pointlike, with emission regions known to be more compact than galaxy scales ($\sim$$100~{\rm kpc}$) at $\gamma$-ray energies, and despite the instrumental point-spread functions (PSFs) of such telescopes being far from being diffraction limited, that is, $\lambda/D<<1$, where $D$ is the telescope diameter, which is typically $\sim$$1~{\rm m}$ \citep{Perlman2015, Steinbring2015}. Even so, for $\alpha=0.67$, the effect should be obvious in a GRB of a fairly modest redshift, say $z=0.10$, emitting photons at 100 MeV, as each wavefront phase dispersion $\Delta \phi_0$ is over 1 radian. Interpreted as an equivalent ray deflection, this would uniformly ``scatter” those photons out over a disk of the same solid angle, complicating identification. And, as such sources are found, Perlman et al. consider the holographic formulation excluded, allowing only the most phase-dispersed wavefronts (lowest $\alpha$) while neglecting in the PSF any less-scattered photon arriving from a given redshift \citep{Ng2022}. Alternatively, Steinbring \cite{Steinbring2015} sums the total, cumulative PSF within the horizon (i.e., for $\Delta\phi\leq 2\pi$) integrating less-scattered photons (assuming these could be affected by blurring due to any higher value of $\alpha$) including those from along the path to that redshift, within the field of view (FoV), in addition to diffraction, and down through to the Planck scale. Averaged this way, and designated $\Phi$, it predicts that sources should be detectable well beyond 10 GeV (not all photons scattered outside 1 radian) %***Author: blurred->scattered ***
and that the general effect is instead degraded localizability (i.e., blurred image cores, with $\alpha$ setting the maximal PSF) falling within a surrounding halo \citep{Steinbring2016}. This effect is independent of $\gamma$-ray pair-production haloes, which might also be spread over $\sim$${\rm Mpc}$ scales for sources with $z<0.5$, as discussed in \cite{Steinbring2015}. Those have been looked for, without success, around low-redshift blazars by stacking together the data of many sources, for example, \citep{Chen2015}. This is also an effect unrelated to the absorption of $\gamma$-rays by the extragalactic background light: foam-induced blurring would only spread out an image of a distant source and so will not reduce the flux of a GRB.

\begin{figure}[H]
%\begin{figure}
\includegraphics[scale=0.3]{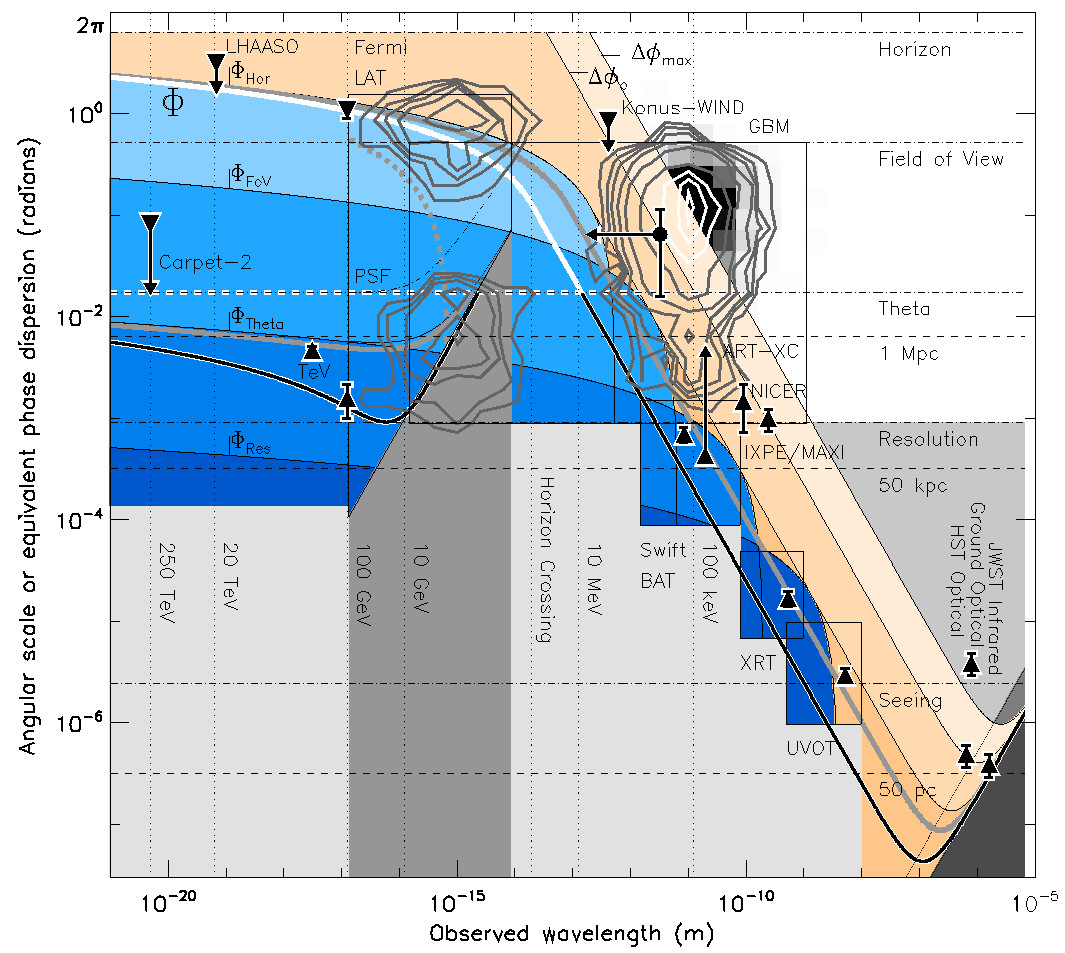}
\includegraphics[scale=0.3]{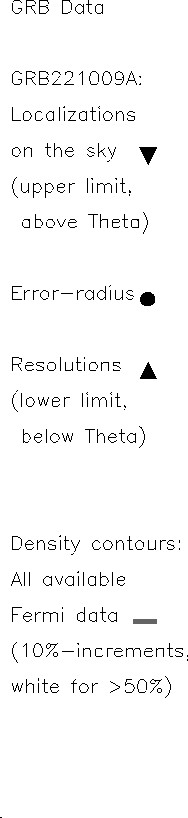}
%\plotone{figure_wide.eps}
\caption{Angular localization (or equivalent wavefront phase dispersion) of GRB221009A: {\emph{ Fermi}} LAT and GBM; {\emph{ Swift}} BAT, XRT and UVOT; or other space- and ground-based facilities. FoVs and energy-sensitivity ranges of instruments are indicated by thin black outlines, and scaled LAT PSF is that measured from AGNs (dot-dashed curve; see Appendix~\ref{app}); gray-shaded regions are instrument resolutions, with increasingly dark shading indicating diffraction: {\emph{ Fermi}} LAT ($D=1.8~{\rm m}$), darker, {\emph{ HST}} ($D=2.4~{\rm m}$) and darkest {\emph{ JWST}} ($D=6.5~{\rm m}$). Maximal limits imposed by foam-induced blurring (orange-shaded regions; $\alpha=0.650$ and $z=4.61$) and combination with instrument viewing angles (blue) are shown; their average $\Phi$ (gray curves, $\alpha=0.667$ and $z=1.41$) and minima (thick white curve above Theta, thick black below; $\alpha=0.735$ and $z=0.151$) follows in Section~\ref{model}; apparent source sizes (dashed horizontal lines, calculated for $z=0.151$) and localization/resolution (horizontal dot-dashed); for details of all other available data, which are shown here as density contours, see Section~\ref{comparison}.}\label{figure_wide}
%\end{figure}
\end{figure}

A special GRB in this regard is the brightest ever observed: GRB221009A. It triggered the \emph{Fermi} observatory Gamma-ray Burst Monitor (GBM) within its FoV of $35^\circ$, which located it in the sky to within a error radius of $3.71^\circ$ (90\% confidence) at peak energy 375~keV~\citep{Veres2023}, and also with the Large Area {Telescope} %MDPI: we deleted [LAT;][], please confirm. ***Author: Okay.***
 \citep{Bissaldi2023}. The latter has a resolution of $5^\circ$ at 30 MeV or 1.5'~at 60 GeV and, together with the GBM, has found over 3390 GRBs (median-$z$: 1.41; highest-$z$: 4.61) from 100 MeV to 100 GeV since its launch in 2008; shown in Figure~\ref{figure_wide} are the  contour plots of all the localization data available from the public archive ({including} %MDPI: foootnote is not allowed, we moved into maintext, please confirm current version; same with other footnote ***Author: Okay, as edited.***
 GRB221009A; this is the full {\emph{ Fermi}} GRB database, as accessed on 1 November 2022 from the High-Energy Astrophysics Science Research {Archive:} %MDPI: Please add the access date (format: Date Month Year), e.g., accessed on 1 January 2020.  ***Author: It is noted; "as accessed on 1 November 2022."***
  \url{https://heasarc.gsfc.nasa.gov/}) plotted at the peak detected wavelength, as scaled by $1/(2\pi{\rm c}\hbar)$. Although GRB-monitoring instruments like the LAT and GBM cannot resolve those sources, the LAT does provide a measure of how far the telescope must slew---called a roll angle---to center the GRB within the instrument FoV, which is something less than the zenith angle to the celestial pole. This automatic repointing is not currently operational, nor was it during the GRB221009A trigger, even though that does not affect the results here. For very high energy sources like GRB221009A, this angle can be several factors larger than the nominal instrumental PSF, making that an underestimate \citep{Ajello2021}. Despite this, once triggered, finding GRB221009A with the LAT can restrict blurring, as photons were detected at 100 GeV within 1 radian for either measure, including at 397.7 GeV \citep{Xia2022}, so not all could have been scattered to the horizon. These two especially high-energy photons are, however, problematic: the first was detected 240 s after the GBM trigger, and so suffers from being in the bad-time-interval (BTI) region, which has compromised utility ({for a discussion of this BTI} data anomaly with {\emph{ Fermi}} LAT/GBM observations of GRB221009A, {see} %MDPI: Please add the access date (format: Date Month Year), e.g., accessed on 1 January 2020.
 \url{https://fermi.gsfc.nasa.gov/ssc/data/analysis/grb221009a.html}). And the second photon, of even higher energy, followed almost 10 h later, which is hard to explain physically with the current GRB theory \citep{Xia2022}. Within the following discussion, these datapoints will be retained as reported, with those caveats.

And with GRB221009A, for the first time, at least one higher-energy photon may be inferred by Carpet-2 via the  cosmic-ray air-shower angle of arrival (AoA) at 251 TeV \citep{Dzhappuev2023}. The reconstructed AoA was at ${\rm R.A.}=289.51^{\circ}$ and ${\rm Dec.}=18.44^{\circ}$, detected 1338 s after triggering by {\emph{ Swift}} and 4536 s after the {\emph{ Fermi}} GBM. Carpet-2 has an all-sky FoV (essentially $2\pi$) and a resolution of $4.7^\circ$ (90\% confidence) indicated by a down-pointing arrow in Figure~\ref{figure_wide}, with the minimum set by the angular distance to the optical transient, fixing it within $1.78^\circ$. This sets a critical angle near $1^\circ$ (about half of the localization accuracy), because if foam-induced blurring is present at this wavelength, it likewise demands here (and at all longer wavelengths) some wavefronts phase-dispersed less: hereafter, to highlight the importance of this benchmark angle, it is called ``Theta''. That phase-error angle also happens to be near the mid-point of the {\emph{ Fermi}} LAT/GBM resolutions, which is nominally the mean PSF of those instruments. For the LAT, it has been measured via images of a large sample of AGNs (see {Appendix~\ref{app}}). Table~\ref{table_limits} lists these limits and the other relevant data: {\emph{ Swift}} satellite detection \citep{Dichiara2023} with the Burst-Alert Telescope (BAT), X-ray Telescope (XRT) and Ultraviolet Optical Telescope (UVOT), as outlined in Figure~\ref{figure_wide}; the black symbols indicate the positional accuracy, both the  roll angle/zenith angle or error radius (above Theta) and the localization/resolution (below Theta). The most stringent of these in each energy range are plotted at 18 TeV by the Large High-Altitude Air-Shower Observatory (LHAASO) \citep{Huang2022}, %[LHAASO;][]
 allowing for only a broad sky localization much like the LAT roll angle, and the space-based instruments Konus-{\emph{WIND}} (KW) and {\emph{ Mikhail } \emph{Pavlinsky}} Astronomical Roentgen Telescope X-ray Concentrator (ART-XC) \citep{Fredericks2023}%[ART-XC;][]
, the Neutron Star Interior Composition Explorer (NICER) \citep{Iwikiri2022}%[NICER;][]
, the Monitor of the All-sky X-ray Imager (MAXI) \citep{Negoro2022}%[MAXI;][]
and the Imaging X-ray Polarimetry Explorer (IXPE) \citep{Negro2022}, %[IXPE;][]
along with imaging from the \emph{Hubble Space Telescope} (HST) \citep{Levan2022a}, %[HST;][]
the {\emph{James} \emph{Webb} \emph{ Space} \emph{ Telescope}} (JWST) \citep{Levan2022b} %[JWST;][]
and a redshift obtained by using ground-based optical telescope spectroscopy \citep{deUgartePostigo2023}. Thus, having located the object in the sky to within a degree in higher-energy $\gamma$-rays than seen before, and then later being able to identify its host galaxy, new restrictions can be placed on the blurring attributable to spacetime foam.

\begin{table}[H]
\caption{{\emph{ Fermi}, \emph{ Swift}} %MDPI: e formated all tables (header, lines, alignment etc.), please check and confirm current versions.
 or other angular limits and localization accuracies for GRB221009A.}
\label{table_limits}
\newcolumntype{C}{>{\centering\arraybackslash}X}
\begin{tabularx}{\textwidth}{CCC}
\toprule
{\textbf{Telescope or Instrument}} & {\textbf{Peak E or} \boldmath{$\lambda$}} & {\textbf{Angle}}\\
\midrule
{{\bf Horizon}} %MDPI: Please add an explanation for bold AND italics in the table footer. If the bold AND italics  is unnecessary, please remove it. 
                             &-                 &$2\pi$\\
\phantom{   }Carpet-2                     &251 TeV           &$1.78^\circ$--$4.7^\circ$\\
\phantom{   }LHAASO VHE                   &18 TeV            &$\leq$$180^\circ$\\
\phantom{   }\emph{ Fermi} LAT (roll angle) &99.3 GeV          &$62.1^\circ$\\
{\bf Field of View}                       &-                 &$35^\circ$\\
\phantom{   }Konus-\emph{WIND}             &3.04 MeV          &$\leq$$48.2^\circ$\\
\phantom{   }\emph{ Fermi} GBM              &375 keV           &$3.71^\circ$\\
{\bf Theta}                               &-                 &$1^\circ$\\
\phantom{   }\emph{ Fermi} LAT (extreme)    &397.7 GeV         &$0.27^\circ$\\
\phantom{   }\emph{ Fermi} LAT (resolution) &99.3 GeV          &$0.09^\circ$\\
\phantom{   }\emph{ Swift} BAT              &146 keV           &2.4$'$\\
\phantom{   }ART-XC                       &4--120 keV         &36$'$\\
\phantom{   }NICER/MAXI                   &13.5 keV          &2.5$'$--10$'$\\
\phantom{   }IXPE                         &5 keV             &$3.4'\pm 1.0'$\\
{\bf Resolution}                          &-                 &1$'$\\
\phantom{   }\emph{ Swift} XRT              &2.3 keV           &3.5$'$\\
\phantom{   }\emph{ Swift} UVOT             &5.25 nm           &0.61$''$\\
{\bf Seeing}                              &-                 &$0.5''$--$1.0''$\\
\phantom{   }Ground-based optical         &800 nm            &0.80''\\
\phantom{   }\emph{ HST}                    &650 nm            &0.10''\\
\phantom{   }\emph{ JWST}                   &$1.65~\upmu{\rm m}$ &0.08''\\
\bottomrule
\end{tabularx}
\end{table}

In light of GRB221009A's high-energy $\gamma$-ray localization, Section~\ref{model} restates the phase dispersion plus diffraction limit $\Phi$ relative to angle Theta and how that scales to the instrumental resolution and FoV, allowing it to be compared to the localization of instruments sensitive to lower energies, that is, for sources with peak energies that occur at longer wavelengths, such as in R-rays and in the optical/near infrared. The approach taken is to attempt falsifying $\Phi$ by looking for any case where a localization (either a roll angle, error radius or a partially resolved image size of a point source) is found to be larger than is expected. The detailed agreement with the {\emph{ Fermi}} and {\emph{ Swift}} AGN and GRB data is discussed in Section~\ref{comparison}, finding a best-fit mid-way value of $\alpha=0.667$ for QG-attributable blurring; the  conclusions follow in Section~\ref{conclusions}.

\section{Method of Finding Quantum-Foam Blurring}\label{model}

The expected wavelength-dependent spread in GRB localization is now considered relative to whether, on average, foam-affected photons should fall inside the horizon (allowing the source to be localizable somewhere in the sky) or within some smaller characteristic angle Theta.

\subsection{Maximal Limit and Average Effect Redward of the Horizon-Crossing Wavelength}

At most, for a wavefront propagating through spacetime foam, blurring accumulated by a co-moving distance $L = ({c/{H_0 q_0^2}})[q_0 z - (1 - q_0)(\sqrt{1 + 2 q_0 z} - 1)]/(1+z)$ is$$\Delta\phi_{\rm max}=2\pi a_0 {l_{\rm P}^{\alpha}\over{\lambda}}\Big{\{}\int_0^z L^{1 - \alpha} {\rm d}z + {{(1-\alpha)c}\over{H_0 q_0}}$$ $$~~~~~~~~~~~~~~~~~~~~~~~\times \int_0^z (1+z) L^{-\alpha} \Big{[}1 - {{1 - q_0}\over{\sqrt{1 + 2 q_0 z}}}\Big{]} {\rm d}z \Big{\}}$$ $$~~~~=\Delta\phi_{\rm los} + \Delta \phi_z=(1+z)\Delta\phi_0, \eqno(2)$$where $q_0={{\Omega_0}/{2}} - {{\Lambda c^2}/{3 H_0^2}}$ is the deceleration ({assuming a standard $\Lambda$CDM cosmology; values of $\Omega_\Lambda=0.7$, $\Omega_{\rm M}=0.3$ and $H_0=70~{\rm km}~{\rm s}^{-1}~{\rm Mpc}^{-1}$ are assumed throughout}) parameter \citep{Steinbring2007}. The effect is stronger in bluer light; here, $\Delta\phi_{\rm los}$ includes waves propagating from points along the line of sight, and $\Delta \phi_z$ are exclusively those redshifted to the observer. Note how the ratio between the greatest and the least-possible effect is always $\Delta \phi_{\rm max} / \Delta \phi_{\rm P}=(1 + z) a_0 (L/l_{\rm P})^{1 - \alpha}$, without dependence on $\lambda$, and where $\Delta \phi_{\rm P} = 2\pi{{l_{\rm P}}\over{\lambda}}$, that is, a minimal perturbation corresponding to the Planck length. And so, as long as no photon is scattered to the horizon, a long exposure produces an image averaging all the detectable phase dispersions, and if those have a distribution with amplitude ${\Delta \phi}~\sigma (\Delta \phi) = 1-A \log({{\Delta \phi}/{\Delta \phi_{\rm P}}})$, this is $${1\over{A}} \int \Delta \phi ~\sigma (\Delta \phi) ~{\rm d}{\Delta \phi} = (1 + z) \Delta \phi_0, \eqno(3)$$for $A = 1/\log{[(1 + z) a_0 (L/l_{\rm P})^{1 - \alpha}]}$, recovering equation 1, and constant for all $\lambda$. In short, although some photons are dispersed by $\Delta\phi_0$ (perhaps right up to the maximum), the average blurring at any wavelength is less than this; it is shown in Figure~\ref{figure_wide} for $\alpha=0.650$ and the median $z=1.41$ as a thick gray curve, where the redward of this ``horizon-crossing'' (vertical dotted line in Figure~\ref{figure_wide}) maintains a simple power-law wavelength dependence.

\subsection{Combined Effect, Scaled between Telescope Field of View and Instrumental Angular Resolution}

To explore wavelengths blueward, consider a point source viewed by a telescope that sees dispersions less than $\theta$, that is, its FoV or a lesser opening angle between that and the instrumental resolution. In Figure~\ref{figure_wide}, an angle Theta of $1^\circ$ is chosen (see Table~\ref{table_limits}). For $\theta \leq 2\pi$ and $A>0$, this implies a PSF mean width \citep{Steinbring2015} $$\Phi = R\Big{(}{{\lambda}\over{D}}\Big{)}^\rho + \int_0^\theta \Delta \phi ~\sigma (\Delta \phi) ~{\rm d}{\Delta \phi} ~~~~~~~~~~~~~~~~~~~~$$ $$~~~~~~~~~ = \Phi_R + \Phi_\theta = A R \Big{(}{\lambda\over{D}}\Big{)}^\rho \Big{[} 1 + \log{\Big{(}{{2\pi l_{\rm P} D^\rho}\over{R \lambda^{\rho+1}}}\Big{)}}\Big{]}~~~~~~~~~~~~~~~~~~~~~~~~~~~~~~~~~~~~~~~~~~~~~$$ $$~~~~~~~~~~~~~~~~~~~~~~~~~~~~~~~ + \theta \Big{\{} 1 + A\Big{[} 1 + \log{\Big{(}{{2\pi l_{\rm P}}\over{\theta \lambda}}\Big{)}}\Big{]}\Big{\}}, \eqno(4)$$where the two components arrive from the integration by parts, that is, splitting the integral above and below $R(\lambda/D)^\rho$. For a telescope of diameter $D$, the {$\Phi_{\rm R}$} %MDPI: please check if R should be in italic as in equation, please check all variables whether they should be unified to keep consistent format. ***Author: Good catch; yes, just this R should be in italics; I don't see others***
 portion includes all the phase dispersions up to its resolution limit, e.g., optical telescopes have $R=1.22$ and $\rho=1$. There is a range of wavelengths where $\Phi$, if present, must be more than the telescope resolution and less than the FoV.

Thus, depending on the outer scale sampled by the instrument, there are four distinct, yet self-similar, degrees of PSF blurring, a shape analogous to an (inverted) wedding cake: a narrow ``base'', i.e., a  sharper PSF core (for those photons falling inside the instrument resolution) and progressively larger layers (for those falling outside its FoV and eventually the horizon itself). The last corresponds to the ``top'' of the cake, which is either the roll angle/zenith angle of the \emph{ Fermi}-LAT or blueward edge of the error radius for the GBM. These sub-regimes are illustrated in Figure~\ref{figure_wide} (shaded darker blue with a narrower field) for $\alpha=0.650$ and $z=4.61$: $\Phi_{\rm Hor}$, $\Phi_{\rm FoV}$, $\Phi_{\rm Theta}$ and $\Phi_{\rm Res}$. Notice that these layers all ``turn over'' in the same sense toward higher energy, implying a smooth transition between them: at the top, limited by the horizon, and at the lower edge by instrument resolution, or---in the middle---Theta.  The lower limit where this rises above the nominal resolution occurs at $\alpha=0.735$, shown as the black-on-white curve. In between, scaling from an upper limit of $\Phi_{\rm Hor}$ to $\Phi_{\rm Theta}$, the last term in Equation (4) would follow the ratio $1+2\pi/(1.22\times\theta)$, where $\theta$ is $1^\circ$, and here it is called $\Phi_{\rm Scaled}$. And so, an interesting result is that the half-way point for $\alpha$ (i.e., just scaled instead by the ratio of the resolution to the horizon) would then happen to occur at $\alpha=0.667$, which is the holographic value favored by QG models. This limit, much like ``seeing'' from the ground, can be looked for along with the implied smooth scaling for the average size of the bluest $\gamma$-ray sources (plotted as a dashed gray curve for the average redshift $z=1.41$).

\section{Analysis and Discussion}\label{comparison}

Armed with an expectation for their foam-blurred PSFs, i.e., $\Phi$, in the appropriate regimes, the full catalog of GRB detections are explored further. Figure~\ref{figure_wide} already shows the result of taking all the available localization data for GRB221009A, including from the {\emph{ Fermi}} and {\emph{ Swift}} databases, indicated by contour plots; now, individual datapoints are inspected, trying to find any instance where it is violated. The approach will be to start at lower energies (where agreement with $\alpha\geq 0.650$ is already established), %***Author: Flipped the equality, as it was also the wrong way round; correct now.***
 working toward higher energies and a sharper resolution, and see if agreement continues, including for a higher range of $\alpha$.

\subsection{Alpha Lower Limit by Comparison with Peak-Energy X-rays}

\begin{figure}[H]
%\hspace{10mm}\plotonesmall{figure_colourbar.eps}\\
\includegraphics[scale=0.3]{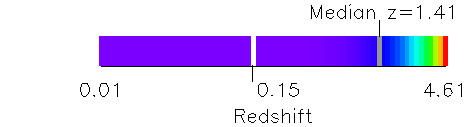}\\
\includegraphics[scale=0.3]{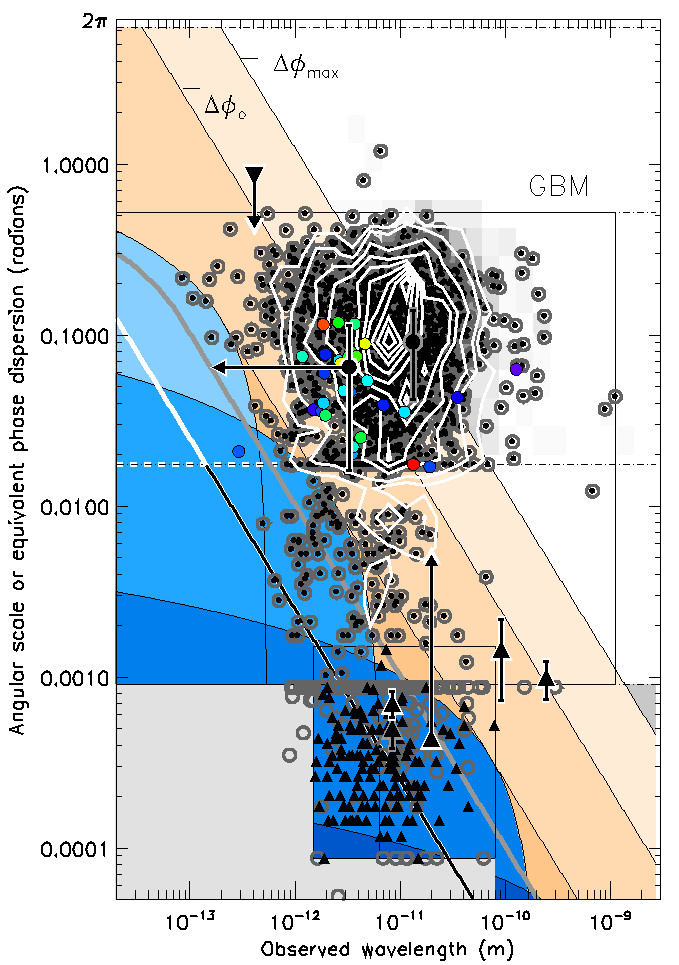}
%\plotonemedium{figure_xray.eps}
\caption{Same as Figure~\ref{figure_wide}, restricted to X-rays: all available {\emph{ Fermi}} GBM peak-energy positional error radii (open gray circles), and {\emph{ Swift}} BAT resolutions (filled up-pointing triangles) for GRBs. Those GBM data beyond its resolution are marked with a black dot, and all angular distances for identified sources are shown color-coded by redshift, with associated color-bar shown above. }\label{figure_xray}
\end{figure}

Consider first only those data at X-ray energies, so sources with peak emission wavelengths long enough that most detected photons would lie redward of the horizon-crossing wavelength. This is shown in Figure~\ref{figure_xray}: all the reported error radii of each GRB localization at peak energy for the {\emph{ Fermi}} GBM instrument (open gray circles, with a central black dot there above the instrumental resolution); all the identified sources are indicated at their mean angular distance from the initial trigger position and shown as  color-coded by the reported redshift, with the associated color-bar shown above.  The density contours (in 10\% increments up to 90\%) help show their skewed distribution with energy. There is a remarkable correspondence between that and the power-law slope of $\Delta\phi_0$ and the lesser, average blurring expected, especially above Theta. Note that the two GBM sources %please confirm intended meaning has been retained ***Author: Okay.***
that are more poorly constrained than $35^\circ$ lie outside the FoV due to the conservative choice of using 99\% containment to define the FoV. Also notice that although some sources fall to the right of the maximal foam-induced blurring case (and so are evidently of a lower peak energy), this does not obviate it as the correct limit.  The prediction is instead that, above Theta (the mean blurring), {\emph{ most}} photons 
%***Author: If an allowed style, please retain my italics on the word "most" to emphasize it.***
should be blurred more than (and so sources should be found to the right, beyond) the gray curve, as is seen. Moreover, this correspondence carries on in the anticipated way below Theta, hemmed by the lower limit set by $\Phi_{\rm Hor}$ (assuming $\alpha=0.650$ and $z=0.151$), which never rises above the observed GRB221009A localization accuracy at these energies. That is true all the way down to the resolution afforded by the GBM (about 1') as well as for the {\emph{ Swift}} BAT resolutions (up-pointing triangles). That would not remain true if $\alpha < 0.650$ (i.e., stronger blurring, shifting these model curves to the right), %***Author: Equality changed to strictly less-than, which is now correct.***
although this was chosen as it is known to agree with the previously established optical/near-infrared AGN/galaxy results, as discussed in Section~\ref{introduction}.

\subsection{Alpha Upper Range by Fit to Mean-Energy Gamma-rays} 

\begin{figure}[H]
%\begin{figure}
%\hspace{10mm}\plotonemedium{figure_colourbar.eps}\\
\includegraphics[scale=0.3]{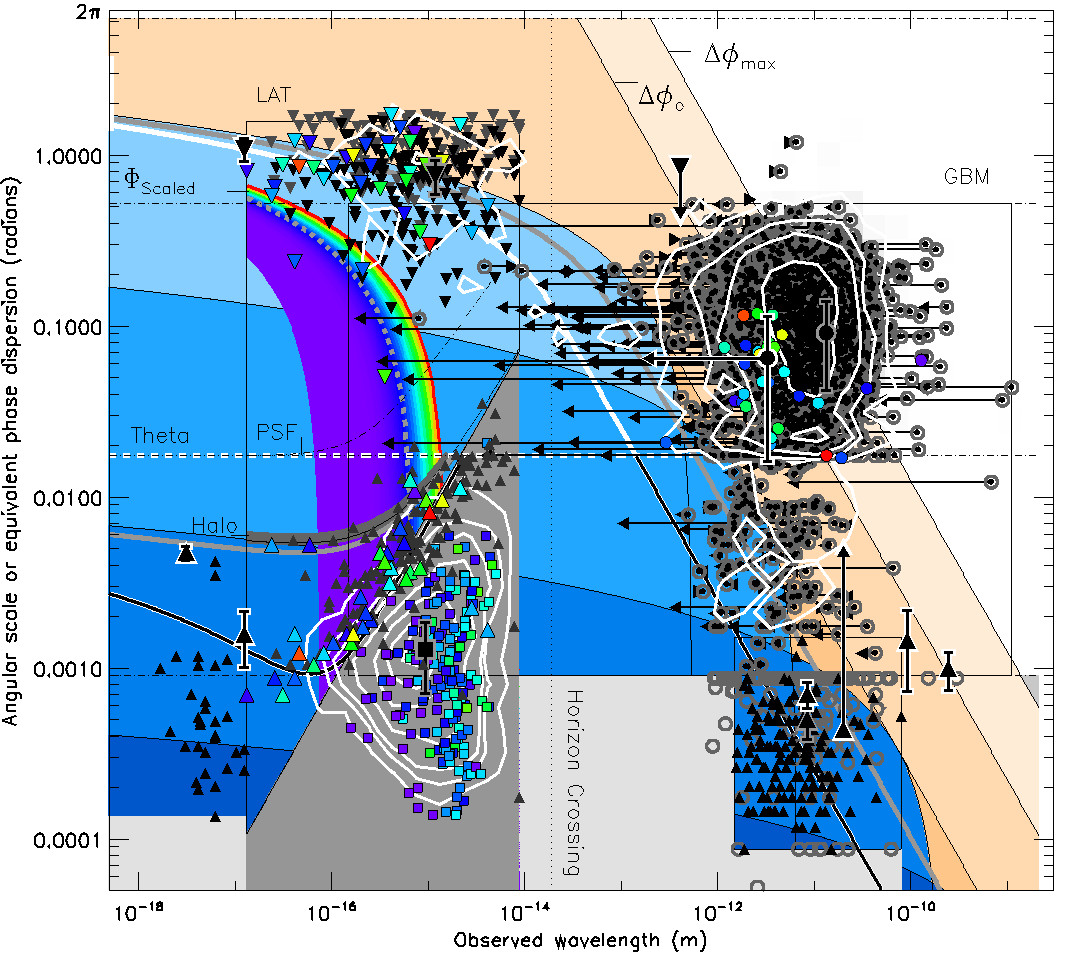}
%\plotone{figure_both.eps}
\caption{Same as Figure~\ref{figure_wide} for $\gamma$-ray and x-ray energies: \emph{ Fermi} LAT roll angles ({black down-pointing triangles)} %MDPI: please check if other colors should be explained.  ***Author: It is correct, and I changed the text below slightly to read "color-coding of *all* symbols is the same as in Figure 2, which is now more clear.***
 and zenith distances (gray); density contours outline AGN point source sizes (see text) with squares indicating those sources of known redshift; color-coding of all symbols is the same as in Figure~\ref{figure_xray}. All GBM-reported GRB peak energies (gray open circles) have arrows (left- or right-pointing triangles) to the mean energy recorded for each source.  Averages within LAT and GBM are indicated by black-on-gray symbols, with one-sigma error bars.}\label{figure_both}
%\end{figure}
\end{figure}

Looking to see if this agreement continues when considering hard X-rays and at $\gamma$-ray energies, those data can be compared to the expected dispersion of photons blueward of the horizon-crossing wavelength. All the {\emph{ Fermi}} LAT and GBM localizations are shown again in Figure~\ref{figure_both}: the roll angles (filled, down-pointing triangles) and zenith angles (same, gray) and the resolutions (up-pointing triangles). The peak energies (GBM) above the resolution limit are indicated by open gray circles, with each reported mean energy (typically less than the peak) indicated by a left-pointing triangle (pointing right in rare cases when equal to the peak). Other higher-energy TeV sources (up-pointing, filled triangles) are from the TeVCat catalog ({for consistency with the \emph{ Fermi} catalog, this is complete as accessed on 1 November 2022} {from} %MDPI: Please add the access date (format: Date Month Year), e.g., accessed on 1 January 2020. ***Author: Yes, it is given as 1 November 2022, so okay now.***
 \url{http://tevcat.uchicago.edu}). The redshift dependence of $\Phi_{\rm Scaled}$ is indicated with the same color-scaling as that in the color-bar of Figure~\ref{figure_xray}. The averages for each are indicated by gray symbols, with one-standard-deviation error bars. The white contours are the 5\%, 10\%, 50\% and 90\% containment of all the LAT plus GBM roll-angle and error-radii data, which now have both their peaks and averages in a combined sample. Notably, the density contours of the combined \emph{ Fermi} LAT roll angles and GBM error radii are matched by the mean $\Phi$ (thick white curve above Theta: $\alpha=0.650$ and $z=0.151$) with a smooth transition blueward to lie within $\Phi_{\rm FoV}$; this gives some confidence that foam-induced blurring is the underlying cause of the broad spread in both of those datasets.

Notice how well the upper limit of the blurring matches the mean distribution of the roll angles (including GRB221009A) in the upper left of Figure~\ref{figure_both} for the LAT data by themselves. Below that, most remarkable is that the model of the foam-induced blurring plus LAT instrument resolution limit (thick gray curve labeled ``Halo'') fits the locus of the poorest-resolved GRBs well. And the  agreement nicely continues to where they turn off in approaching angle Theta at the lowest LAT energies. In other words, this explains the ``tailing off'' of the largest image sizes at the longest wavelengths detectable with the LAT, despite what might naively be expected to become worse there due to the effects of diffraction (the slope of the gray-shaded region). That upper edge of the GRB datapoints also closely matches the upper extent of the AGN-size samples from the LAT, taken from the Fourth point source catalog (Abdollahi et al., 2020; 4050~samples, mean-$z$: 0.967) and shown as white density contours (1\%, 5\%, 10\%, 25\%, 50\%, 75\% and 90\%; and, where available, filled squares color-coded by redshift using the scale in Figure~\ref{figure_xray}) from which the telescope instrumental PSF is reported (see Appendix~\ref{app}). Equation~(4) anticipates this behavior, because if these are true point sources of a given redshift blurred only by foam, the last term in that equation allows for each being able to accumulate any image size {\emph{ smaller}} %***Author: If the style is alllowed, I'd prefer this word in italics to help emphasize it.*** 
than Theta, down to the resolution limit---even though that may be less than the mean resolution at that energy; the LAT images are not limited by diffraction. This instead implies that foam-induced blurring sets that lower limit, a resolution floor indicated here by the darkest blue shading. Indeed, for the highest energies detectable with the LAT (shortest wavelengths), the weakest possible $\Phi_{\rm Theta}$ (black-on-white curve, $\alpha=0.735$ and $z=0.151$) is consistent with the localization of GRB221009A. And, in between, the average redshift (dashed curve) agrees with the mean energies of all the LAT and GBM sources as well, that is, either the down- and left-pointing arrows for any of those sources could lie anywhere within the sensitivity of the  LAT and GBM, but blurred by holographic QG foam ($\alpha=0.667$) none of those should stray blueward of this smooth, redshift-scaled demarcation (color-coded in the same way as in Figure~\ref{figure_xray}). That does seem to be the case, judging that one sigma from the mean angular localization of the sample is a fair estimate of uncertainty in these reported data, accommodating two possible outliers (one dark blue and one green down-pointing triangle) in the LAT GRB roll-angle sample. Overall, Equation (4) shows good agreement with all the data at these energies.

\section{Summary and Conclusions}\label{conclusions}

Quantum-foam-induced blurring can explain the broad multi-degree angular spread in the localization of GRBs, when detecting those objects in the sky in high-energy $\gamma$-rays and X-rays, and yet still allow these to be identified at %attention AE: please review spelling of unit of measure; ***Author-response: fixed***
sub-arcsecond scales within the source galaxy by its longer wavelength galaxian emission, that is, via later optical/near-infrared spectroscopy. This is an improvement on earlier descriptions of foam-induced blurring, now dealing with the resolution and angular spread of pointlike sources, and so with the effect of photons falling either outside the telescope FoV or within the resolution of an instrument. A characteristic angle ``Theta'' has been introduced, which is incorporated into $\Phi$, and is analogous to the image quality of ground-based telescopes due to uncorrected atmospheric turbulence. That is a useful concept, setting the mean angular size $\Phi_{\rm Theta}$ of any GRB to be under $1^\circ$ near 250 TeV, agreeing with the highest-energy localization of GRB221009A, although it is still consistent with $\alpha\geq 0.65$ and so with previous {\emph{ HST}} imaging of galaxies, AGNs and all other available GRB localizations.

Twofold progress over previous investigations of foam-induced blurring of GRBs has been made: First, beyond being in accordance with the highest-energy Carpet-2 observation of GRB221009A, this is in agreement with the in-flight-measured PSF of the {\emph{ Fermi}} LAT and an upper GRB size (a ``Halo'' angle) bounded at low energy by Theta. That is of interest to GRB and AGN studies, implying that the PSFs used to measure their sizes have neglected the effect of foam-induced blurring: so, in essence, this is akin to having ignored atmospheric seeing in measuring the sizes of ground-based optical/near-infrared galaxies, warranting further scrutiny. And, secondly, this investigation also moves searches for foam-induced blurring forward by providing more than simply a lower limit of the parameter $\alpha$. It instead implies the observed effect falls below that and within an upper range, i.e., \mbox{$0.650\leq \alpha\leq 0.735$}, which also agrees with an average, scaled value of 0.667. That is of broader interest to physics in general, as the further implication is that high-energy observations of GRB221009A may constitute the first experimental evidence favoring a holographic formulation of QG. 

%%%%%%%%%%%%%%%%%%%%%%%%%%%%%%%%%%%%%%%%%%
\vspace{6pt} 

\funding{{This research received no external funding.}} %MDPI: Please add: ``This research received no external funding'' or ``This research was funded by NAME OF FUNDER grant number XXX.'' and  and ``The APC was funded by XXX''. Check carefully that the details given are accurate and use the standard spelling of funding agency names at \url{https://search.crossref.org/funding}, any errors may affect your future funding.}
\dataavailability{All data discussed herein are freely available via public archives using the links indicated in the text, and their references. Those data are also all downloadable from \url{https://github.com/ericsteinbring/Special-Blurring} in formatted data tables, together with the analysis code, written in IDL. In its default settings, that code generates the figures shown here.} %***Author: I've moved the data-availability statement from the acknowledgements to here.***
%MDPI: We encourage all authors of articles published in MDPI journals to share their research data. In this section, please provide details regarding where data supporting reported results can be found, including links to publicly archived datasets analyzed or generated during the study. Where no new data were created, or where data is unavailable due to privacy or ethical restrictions, a statement is still required. Suggested Data Availability Statements are available in section ``MDPI Research Data Policies'' at \url{https://www.mdpi.com/ethics}.
%normally, the DAS for research articles can state "Data are contained within the article." or "Data are contained within the article and supplementary materials." Please refer to the complete guideline at https://www.mdpi.com/ethics#_bookmark21} 

% Only for journal Nursing Reports
%\publicinvolvement{Please describe how the public (patients, consumers, carers) were involved in the research. Consider reporting against the GRIPP2 (Guidance for Reporting Involvement of Patients and the Public) checklist. If the public were not involved in any aspect of the research add: ``No public involvement in any aspect of this research''.}

% Only for journal Nursing Reports
%\guidelinesstandards{Please add a statement indicating which reporting guideline was used when drafting the report. For example, ``This manuscript was drafted against the XXX (the full name of reporting guidelines and citation) for XXX (type of research) research''. A complete list of reporting guidelines can be accessed via the equator network: \url{https://www.equator-network.org/}.}

\acknowledgments{I thank the anonymous referees for many helpful comments and suggestions that improved the final manuscript.}

\conflictsofinterest{{The author declares no conflict of interest.}} %MDPI: Declare conflicts of interest or state ``The authors declare no conflict of interest.'' Authors must identify and declare any personal circumstances or interest that may be perceived as inappropriately influencing the representation or interpretation of reported research results. Any role of the funders in the design of the study; in the collection, analyses or interpretation of data; in the writing of the manuscript; or in the decision to publish the results must be declared in this section. If there is no role, please state ``The funders had no role in the design of the study; in the collection, analyses, or interpretation of data; in the writing of the manuscript; or in the decision to publish the results''.} 

%%%%%%%%%%%%%%%%%%%%%%%%%%%%%%%%%%%%%%%%%%
%% Optional
%\sampleavailability{Samples of the compounds ... are available from the authors.}

%% Only for journal Encyclopedia
%\entrylink{The Link to this entry published on the encyclopedia platform.}

%\abbreviations{Abbreviations}{
%The following abbreviations are used in this manuscript:\\
%
%\noindent 
%\begin{tabular}{@{}ll}
%MDPI & Multidisciplinary Digital Publishing Institute\\
%DOAJ & Directory of open access journals\\
%TLA & Three letter acronym\\
%LD & Linear dichroism
%\end{tabular}
%}

%%%%%%%%%%%%%%%%%%%%%%%%%%%%%%%%%%%%%%%%%%
%% Optional
\appendixtitles{yes} % Leave argument "no" if all appendix headings stay EMPTY (then no dot is printed after "Appendix A"). If the appendix sections contain a heading then change the argument to "yes".
\appendixstart
\appendix
\section[\appendixname~\thesection]{The In-Flight-Measured \emph{Fermi} LAT Point-Spread Function}\label{app}
%\subsection[\appendixname~\thesubsection]{}
The LAT is a pair-production imaging instrument, sensitive to $\gamma$-rays through the resultant electron/positron tracks. Scintillometers covering the array are used to reject background events and a calorimeter allows for the recovery of incident $\gamma$-ray energy. The area of sensed $\gamma$-rays is one sample of the measured resolution (which is much larger than its diffraction limit) and the overall average fit of those (at a given energy) can be considered the instrumental PSF of the telescope. The post-launch PSF was measured with a large sample of AGNs, finding the scaling relation $${\rm PSF}\propto\sqrt{[(C_0 E/100)^{-\rho}]^2 + C_1^2},$$where $E$ is the energy in MeV; the fitting constants are $C_0=3.5^\circ$ and $C_1=0.15^\circ$; and $\rho=0.8$ is the power-law slope \citep{Ackermann2013}. 

%\section[\appendixname~\thesection]{}
%All appendix sections must be cited in the main text. In the appendices, Figures, Tables, etc. should be labeled, starting with ``A''---e.g., Figure A1, Figure A2, etc.

%%%%%%%%%%%%%%%%%%%%%%%%%%%%%%%%%%%%%%%%%%
\begin{adjustwidth}{-\extralength}{0cm}

\reftitle{References}

\PublishersNote{}
\end{adjustwidth}

\begin{thebibliography}{999}
\bibitem[Wheeler(1957)]{Wheeler1957} Wheeler, J.A. On the Nature of Quantum Geometrodynamics. {\it Annals of Physics} {\bf 1957}, {\it 2}, 604-614.
\bibitem[Carlip(2023)]{Carlip2023} Carlip, S. Spacetime foam: a review. {\it Reports on Progress in Physics} {\bf 2023}, {\it 86}, 066001.
\bibitem[Lieu \& Hillman(2003)]{Lieu2003} Lieu, R.; Hillman, L.W. The Phase Coherence of Light from Extragalactic Sources—Direct Evidence Against First Order Planck Scale Fluctuations in Time and Space. {\it Astrophysical Journal} {\bf 2003}, {\it 585}, L77-L80.
\bibitem[Ng, Christiansen, \& van Dam(2003)]{Ng2003} Ng, Y.J.; Christiansen, W.A.; van Dam, H. Probing Planck-scale Physics with Extragalacic Sources? {\it Astrophysical Journal}, {\bf 2003}, {\it 591}, L87-L90.
\bibitem[Ragazzoni, Turatto, \& Gaessler(2003)]{Ragazzoni2003} Ragazzoni, R.; Turatto, M.; Gaessler, W. The Lack of Observational Evidence for the Quantum Structure of Spacetime at Planck Scales. {\it Astrophysical Journal} {\bf 2003}, {\it 587}, L1-L4.
\bibitem[Steinbring(2007)]{Steinbring2007} Steinbring, E. Are High-Redshift Quasars Blurry? {\it Astrophyscial Journal} {\bf 2007}, {\it 655}, 714-717.
\bibitem[Christiansen et al.(2011)]{Christiansen2011} Christiansen, W.A.; Ng, Y.J.; Floyd, D.J.E.; Perlman, E.S. Limits on spacetime foam. {\it Physical Review D} {\bf 2011}, {\it 83}, 84003.
\bibitem[Perlman et al.(2011)]{Perlman2011} Perlman, E.S.; Ng, Y.J.; Floyd, D.J.E.; Christiansen, W.A. Using Observations of Distant Quasars to Constrain Quantum Gravity. {\it Astronomy and Astrophysics} {\bf 2011}, {\it 535}, L9.
\bibitem[Tamburini et al.(2011)]{Tamburini2011} Tamburini, F.; Cuofano, C.; Della Valle, M.; Gilmozzi, R. No quantum gravity signature from the farthest quasars. {\it Astronomy and Astrophysics} {\bf 2011}, {\it 533}, A71-A75.
\bibitem[Cooke et al.(2020)]{Cooke2020} Cooke, R.; Welsh, L.; Fumagalli, M.; Pettini, M. A limit on Planck-scale froth with ESPRESSO. {\it Monthly Notices of the Royal Astronomical Society} {\bf 2020}, {\it 494}, 4884-4890.
\bibitem[Amelino-Camelia et al.(1998)]{Amelino-Camelia1998} Amelino-Camelia, G.; Ellis, J.; Mavromatos, N.E.; Nanopoulos, D.V.; Sarkar, S. Tests of quantum gravity from observations of $\gamma$-ray bursts. {\it Nature} {\bf 1998}, {\it 393}, 763-765.
\bibitem[Liu, Zhang \& Meng(2022)]{Liu2022}Liu, Z.-K.; Zhang, B.-B.; Meng, Y.-Z. Spectral Lag Transition of 32 Fermi Gamma-Ray Bursts and Their Application on Constraining Lorentz Invariance Violation. {\it Astrophysical Journal} {\bf 2022}, {\it 935}, 79, 8pp.
\bibitem[Perlman et al.(2015)]{Perlman2015} Perlman, E.S.; Rappaport, S.A.; Christiansen, W.A.; Ng, Y.J.; DeVore, J.; Pooley, D. New Constraints on Quantum Gravity from X-Ray and Gamma-Ray Observations {\it Astrophysical Journal}  {\bf 2015}, {\it 805}, 10-20.
\bibitem[Steinbring(2015)]{Steinbring2015} Steinbring, E. Detectability of Planck-Scale-Induced Blurring with Gamma-Ray Bursts {\it Astrophysical Journal} {\bf 2015}, {\it 802}, 38-43.
\bibitem[Ng \& Perlman(2022)]{Ng2022} Ng, Y.J.; Perlman, E. Probing Spacetime Foam with Extragalactic Sources of
High-Energy Photons {\it Universe} {\bf 2022}, {\it 8}, 382, 11pp.
\bibitem[Steinbring(2016)]{Steinbring2016} Steinbring, E. Limits to Seeing High-Redshift Galaxies Due to Planck-Scale-Induced Blurring {\it International Astronomical Union Conference Series} {\bf 2016}, {\it 319}, 54.
\bibitem[Chen et al.(2015)]{Chen2015} Chen, W.; Buckley, J.H.; Ferrer, F. Search for GeV $\gamma$-Ray Pair Halos Around Low Redshift Blazars {\it Physical Review Letters} {\bf 2015}, {\it 115}, 211103.
\bibitem[Veres et al.(2023)]{Veres2023} Veres, P.; Burns, E.; Bissaldi, E.; Lesage, S.; Roberts, O. GRB 221009A: Fermi GBM detection of an extraordinarily bright GRB. {\it GRB Coordinates Network, Circular Service}, {\bf 2022}, 32636.
\bibitem[Bissaldi et al.(2023)]{Bissaldi2023} Bissaldi, E.; Omodei, N.; Kerr, M. GRB 221009A or Swift J1913.1+1946: Fermi-LAT detection. {\it GRB Coordinates Network, Circular Service} {\bf 2022}, 32637.
\bibitem[Ajello et al.(2021)]{Ajello2021} Ajello, M.; Atwood, W.B.; Axelsson, M.; Bagagli, R.; Bagni, M.; Baldini, L.; Bastieri, D.; Bellardi, F.; Bellazzini, R.; Bissaldi, E.; et al. Fermi Large Area Telescope Performance after 10 Years of Operation. {\it Astrophysical Journal Supplement Series} {\bf 2021}, {\it 256}, 12, 25pp.
\bibitem[Xia et al.(2022)]{Xia2022} Xia, Z.-Q.; Wang, Y.; Yuan, Q.; Fan, Y.-Z. GRB 221009A: a 397.7 GeV photon observed by Fermi-LAT at 0.4 day after the GBM trigger. {\it GRB Coordinates Network, Circular Service} {\bf 2022}, 32748.
\bibitem[Dzhappuev et al.(2023)]{Dzhappuev2023} Dzhappuev, D.D.; Afashokov, YuZ, Dzaparova, I.M.; Dzhatdoev, T.A.; Gorbacheva, A.; Karpikov, I.S.; Khadzhiev, M.M.; Klimenko, N.F.; Kudzhaev, A.U.; Kurenya, A.N.; et al. Swift J1913.1+1946/GRB 221009A: detection of a 250-TeV photon-like air shower by Carpet-2. {\it The Astronomer's Telegram} {\bf 2023}, 15669.
\bibitem[Dichiara et al.(2023)]{Dichiara2023} Dichiara, S.; Gropp, J.D.; Kennea, J.A.; Kuin, N.P.M.; Lien, A.Y.; Marshall, F.E.; Tohuvavohu, A.; Williams, M.A. Swift J1913.1+1946 a new bright hard X-ray and optical transient. {\it The Astronomer's Telegram} {\bf 2023}, 15650.
\bibitem[Huang et al.(2022)]{Huang2022} Huang, Y.; Shicong, H.; Chen, S.; Zha, M.; Liu, C.; Yao, Z.; Cao, Z. LHAASO observed GRB 221009A with more than 5000 VHE photons up to around 18 TeV. {\it GRB Coordinates Network, Circular Service} {\bf 2022}, 32677.
\bibitem[Fredericks et al.(2023)]{Fredericks2023} Fredericks, D.; Svinkin, D.; Lysenko, A.L.; Molkov, S.; Tsvetkova, A.; Ulanov, M.; Ridnaia, A.; Lutovinov, A.A.; Lapshov, I.; Tkachenko, A.;  et al. Properties of the Extremely Energetic GRB 221009A from Konus-WIND and SRG/ART-XC Observations. {\it Astrophysical Journal} {\bf 2023}, {\it 949}, L7, 11pp.
\bibitem[Iwikiri et al.(2022)]{Iwikiri2022} Iwikiri, W.; Jaisawal, G.K.; Younes, G.; Wadiasingh, Z.; Guillot, S.; Gendreau, K.C.; Arzoumanian, Z.; Ferrara, E.C.; Mihara, T.; Pasham, D.; et al. GRB 221009A: NICER follow-up observations {\it The Astronomer's Telegram} {\bf 2022}, 15664.
\bibitem[Negoro et al.(2022)]{Negoro2022} Negoro, H.; Nakajima, M.; Kobayashi, K.; Tanaka, M.; Soejima, Y.; Mihara, T.; Kawamuro, T.; Yamada, S.; Tamagawa, T.; Matsuoka, M. et al. MAXI/GSC detection of the new X-ray transient Swift J1913.1+1946. {\it The Astronomer's Telegram} {\bf 2022}, 15651.
\bibitem[Negro et al.(2022)]{Negro2022} Negro, M.; Manfreda, A.; Omodei, N.; Muleriet, F. GRB 221009A: IXPE preliminary upper limits to X-ray polarization {\it The Astronomer's Telegram} {\bf 2022}, 15678.
\bibitem[Levan et al.(2022a)]{Levan2022a} Levan, A.J.; Barclay, T.; Bhirombhakdi, K.; Burns, E.; Cenko, S.B.; Chrimes, A.A.; D’Avanzo, P.; D’Elia, V.; Della Valle, M.; de Ugarte Postigo, A.; et al. GRB 221009A: Hubble Space Telescope observations. {\it GRB Coordinates Network, Circular Service} {\bf 2022}, 32921.
\bibitem[Levan et al.(2022b)]{Levan2022b} Levan, A.J.; Barclay, T.; Burns, B.; Cenko, S.B.; Chrimes, A.A.; D’Avanzo, P.; D’Elia, V.; Della Valle, M.; de Ugarte Postigo, A.; Fong, W.; et al. GRB 221009A: James Webb Space Telescope Observations. {\it GRB Coordinates Network, Circular Service} {\bf 2022}, 32821.
\bibitem[de Ugarte Postigo et al.(2023)]{deUgartePostigo2023} de Ugarte Postigo, A.; Izzo, L.; Pugliese, G.;  Xu, D.; Schneider, B.; Fynbo, J.P.U.; Tanvir, N.R.; Malesani, D.B.; Saccardi, A.; Kann, D.A.; et al. GRB 221009A: Redshift from X-shooter/VLT. {\it GRB Coordinates Network, Circular Service} {\bf 2023}, 32648.
\bibitem[Ackermann et al.(2013)]{Ackermann2013} Ackermann, M.; Ajello, M.; Allafort, A.; Asano, K.; Atwood, W.B.; Baldini, L.; Ballet, J.; Barbiellini, G.; Bastier, D.; Bechtol, K.; et al. Determination of the Point-Spread Function for the {\it Fermi} Large Area Telescope From On-Orbit Data and Limits on Pair Halos of Active Galactic Nuclei. {\it Astrophysical Journal} {\bf 2013} {\it 765}, 54, 19pp.
%MDPI: please add et al after first 10 author names, please check all et al in references list; same with other highlight.
%MDPI: Please provide more information about the article type, such as book (please provide the name and location of the publisher); online resource (please provide the URL of the website and the date it was accessed (Date Month Year)); or journal article (please provide the name of the journal, the year and volume in which it was published, and the page number). Please refer to https://www.mdpi.com/authors/references for full reference formatting guides; same with other highlight.
%*** Author: Two references were not used in the text, and so I deleted them.***
\end{thebibliography}
\end{document}